\documentclass[aps,twocolumn,prb,amsmath,amssymb]{revtex4-1}

\usepackage{amsmath}
\usepackage{amssymb}
\usepackage[varg]{txfonts}
\usepackage{color}
\usepackage[colorlinks]{hyperref}
\usepackage{tikz}
\usetikzlibrary{shapes}
\usepackage{xcolor}

\usepackage{graphicx}% Include figure files
\usepackage{bm}% bold math

\newcommand{\RNum}[1]{\uppercase\expandafter{\romannumeral #1\relax}}

\def\XXint#1#2#3{{\setbox0=\hbox{$#1{#2#3}{\int}$}
    \vcenter{\hbox{$#2#3$}}\kern-.5\wd0}}

\def\be{\begin{equation}}
\def\ee{\end{equation}}

\def\bi{\begin{itemize}}
    \def\ei{\end{itemize}}
\def\bn{\begin{enumerate}}
    \def\en{\end{enumerate}}
\def\bea{\begin{eqnarray}}
\def\eea{\end{eqnarray}}
\newcommand{\bpm}{\begin{pmatrix}}
    \newcommand{\epm}{\end{pmatrix}}

\def\ba{\begin{array}}
    \def\ea{\end{array}}
\def\bd{\begin{displaymath}}
\def\ed{\end{displaymath}}

\renewcommand{\imath}{\hspace{1pt}\mathrm{i}\hspace{1pt}}

\begin{document}
\title{Stoner ferromagnetism, correlated metal and thermoelectricity in partially flat-band materials}
\author{Leyla Majidi}
\affiliation{Department of Physics, Sharif University of Technology, Tehran 14588-89694, Iran}

\author{Abolhassan Vaezi}
\email{vaezi@sharif.edu}
\affiliation{Department of Physics, Sharif University of Technology, Tehran 14588-89694, Iran}
\date{\today}

\author{Mehdi Kargarian}
\email{kargarian@sharif.edu}
\affiliation{Department of Physics, Sharif University of Technology, Tehran 14588-89694, Iran}

\begin{abstract}
Recent discovery of correlated electronic phases in twisted heterostructures raised a surge of interests in studying models and materials with flat bands where the electronic excitations are nearly dispersionless in momentum space. As such, the kinetic energy is quenched and the correlations are enhanced, giving rise to a plethora of unusual magnetic, superconducting and transport behaviors. Finding materials whose energy bands are completely flat is rather challenging, yet those whose dispersion is flat only in a portion of the momentum space might be more accessible in material search. In this work, we propose a partially flat-band system on a square lattice. Using the Hubbard model, it is demonstrated that the suppression of the electronic kinetic energy in the flat portion of the band dispersion drives the system to Stoner ferromagnetism even at very weak interactions, i.e., much smaller than the bandwidth, with significantly enhanced Curie temperature. While the low-energy magnon modes are well defined collective excitations, flat magnon bands can be observed at high energies. We show that the strong interaction leads to reduction of the flat portion of the magnon band. However, tuning the chemical potential at a strong interaction regime may lead to spin density wave at finite wave vectors. Then, focusing on the non-magnetic correlated phase and using dynamical mean-field theory, we demonstrate the appearance of a flat-band induced sharp peak in the density of states in addition to the correlation-induced Mott bands. Furthermore, the large seebeck coefficient and the figure of merit of the proposed partially flat-band model, compared to symmetric regular band models, put them in the category of efficient thermoelectric materials.

\end{abstract}

\maketitle

\section{Introduction}

Over the past decade, we have witnessed a growing interest for the physics in flat-band systems. The wealth and fascinating physics that take place in these systems motivate the search for efficient procedures and strategies for flat-band engineering. Electronic systems with flat bands are predicted to be a fertile ground for hosting emergent phenomena including unconventional magnetism and superconductivity, benefiting from the quenched kinetic energy. Since 1986, original theoretical papers predicted that special lattices, exemplified by Dice's~\cite{Dice}, Lieb's~\cite{lieb}, Mielke's~\cite{mielke_1,mielke_2}, and Tasaki's~\cite{Tasaki} models, have the peculiar property that one or more spectral bands are strictly flat or independent of momentum in the tight binding approximation, arising from either internal symmetries or fine-tuned coupling.

The field of flat-band physics has been recently invigorated by the experimental identification of flat electronic bands in a variety of settings ranging from electronic systems such as twisted bilayer graphene~\cite{TBG_exp}, kagome materials~\cite{kagomeh_exp1,kagomeh_exp2,kagomeh_exp3}, and engineered atomic lattices~\cite{lieb_exp}, to cold atoms in optical lattices~\cite{optical_lattice} and photonic devices~\cite{photonic}. The discovery of superconductivity~\cite{TBG_S}, magnetism~\cite{TBG_F} and insulating behavoir ~\cite{TBG_I} in twisted graphene bilayers (near a magic angle $1.1^{\circ}$) has opened new perspectives in the study of twisted van der Waals heterostructures~\cite{WSe2,WSe22,WS2}. However, these intriguing phenomena are mainly limited to samples with twisted angle near the magic angle, which quickly diminish when the twisted angle deviates slightly due to the prompt increase of bandwidth. Thereupon, among the solid state systems, kagome lattices have become the most promising candidates for flat bands in which their electronic spectrum presents both of Dirac and flat bands~\cite{kagomeh_exp1,kagomeh_exp2,kagomeh_exp3}.

Proposals for the flat-band magnetism and superconductivity have focused on multi-band systems, where one of the multi-bands is flat while others are dispersive. Since perfectly flat bands are not stable against generic perturbations, which typically induce nonzero dispersion, the definition of flat bands is broadened to include partially flat bands that have vanishing dispersion only along particular directions or in the vicinity of special Brillouin zone points~\cite{PFB0,PFB1,PFB2,PFB3}. Partially flat-band materials have been realized in organic $\tau$-type conductors~\cite{organic1,organic2,PFB0}, in-organic materials such as ruthenate superconductors~\cite{ruthen1,ruthen2}, some iron-chalcogenides~\cite{iron1,iron2}, twisted van-der Waals heterostructures such as twisted bilayer graphene~\cite{TBG_exp}, twisted WSe$_2$~\cite{WSe2,WSe22} and twisted WSe$_2$/WS$_2$ heterostructure~\cite{WS2}. Alongside, more recent papers have prepared catalogues of high-quality flat- and partially flat-band materials in two-dimensional systems, especially van der Waals materials from the in-organic crystal structure database~\cite{inorganic}, stoichiometric materials~\cite{stoichiometric}, and simple tight-binding models~\cite{tbmodels}. Presence of a Mott insulating state near the half-filling of the flat region in momentum space and significant enhancement of superconducting transition temperatures~\cite{PFB3,Vaezi:CM2023} as well as triplet superconductivity~\cite{PFB4} indicate that unusual correlation physics can indeed occur in partially flat-band systems.

Motivated by this, here we explore a different flat-band model: one-band system with partially flat dispersion. First, we look into magnetism in the partially flat-band model and try to evaluate the connection between the flat electronic bands and itinerant ferromagnetic order, and the role of magnetic fluctuations using the Hubbard model. Second, we consider the correlated metallic phase of the proposed model and examine the thermoelectric properties in the absence of ferromagnetic instability. Due to the existence of partially flat area, correlation effects are pronounced and we expect to encounter evidence of strong correlation physics despite only weak interactions. In particular, very weak Hubbard interactions promote the formation of Stoner ferromagnetism with amplified transition temperature of order $T_C\simeq 2000^{\circ}-4000^{\circ}$ Kelvin, in comparison with that of other ferromagnets.

Considering the role of quantum magnetic fluctuations, we reveal the unusual features of the magnetic excitations by demonstrating the correlation-dependent high-energy flat magnon bands, connected to a low-energy dispersive part around the $\Gamma$ point. Moreover, going away from the flat portion leads to the appearance of a complex spin density wave at finite wave vectors in addition to the ferromagnetic order at the $\Gamma$ point. These key questions can be addressed using inelastic neutron scattering measurements to probe magnetic excitations and determine their coupling to electronic bands. The observation of the flat magnon band in a ferromagnetic metal, similar to their electronic counterparts~\cite{kagomeh_exp1}, is rare: flat-band magnon modes are only reported in the insulating kagome ferromagnet Cu~\cite{kagomecu}, where a dispersive excitation extending from zero energy and momentum transfer up to $\hbar\omega\simeq 1.8$ meV is connected to a flat excitation, and very recently in ferromagnet kagome metals TbMn$_6$Sn$_6$~\cite{kagomeTb}.

Meanwhile, there is a great interest in finding/designing materials possessing large thermoelectricity to be used for converting heat to electricity, or alternatively be used for refrigeration. Previous efforts were mainly focused on reducing the lattice contributions to the thermal conductivity by superstructures or nanostructures~\cite{thermo1}. While the discovery of large Seebeck coefficient, which measures the voltage drop across a system due to a temperature gradient, in transition-metal compounds such as FeSi~\cite{FeSi}, Na$_x$CoO$_2$~\cite{NaxCoO}, and FeSb~\cite{FeSb} with the latter displaying an astonishing response of up to $S=-45$ mV/K at $12^\circ$ K, have shown that strong interactions lead to a large Seebeck coefficient in metals and transition-metal oxides.

To this end, making use of the dynamical mean-field theory, we explore the thermoelectric properties of the proposed structure by evaluating the seebeck coefficient and figure of merit as well as the density of states in non-magnetic phase. The system undergoes Mott transition for the Hubbard interactions $U$ exceeding the bandwidth $W$. Interestingly, an additional sharp peak appears in the density of states manifesting the effect of the flat portion of the electronic spectrum. We find that the seebeck coefficient has a nonmonotonic temperature dependence, becomes saturated at high temperatures, and changes sign as a function of temperature. The sign change depends on the band filling and occurs at higher $U/W$ ratios, with decreasing the band filling. Moreover, the proposed structure has a large figure of merit, in comparison with that of a regular band model, specially at very weak interaction regime manifesting the flatness of the electronic spectrum.

The paper is organized as follows. In Sec. \ref{model}, we introduce the partially flat-band model and characterise the magnetic phase diagram. The magnetic collective modes are studied in Sec. \ref{collective}. Section \ref{thermoelectric} is devoted to investigating the thermoelectric properties of the proposed partially flat-band model in the correlated metallic phase by employing the dynamical mean-field theory to address the density of states, seebeck coefficient and figure of merit. Finally, a brief summary of results is given in Sec. \ref{conclusion}.

\section{Model and magnetic phase diagram}\label{model}
\begin{figure}[h]
\begin{center}
\includegraphics[width=3.4in]{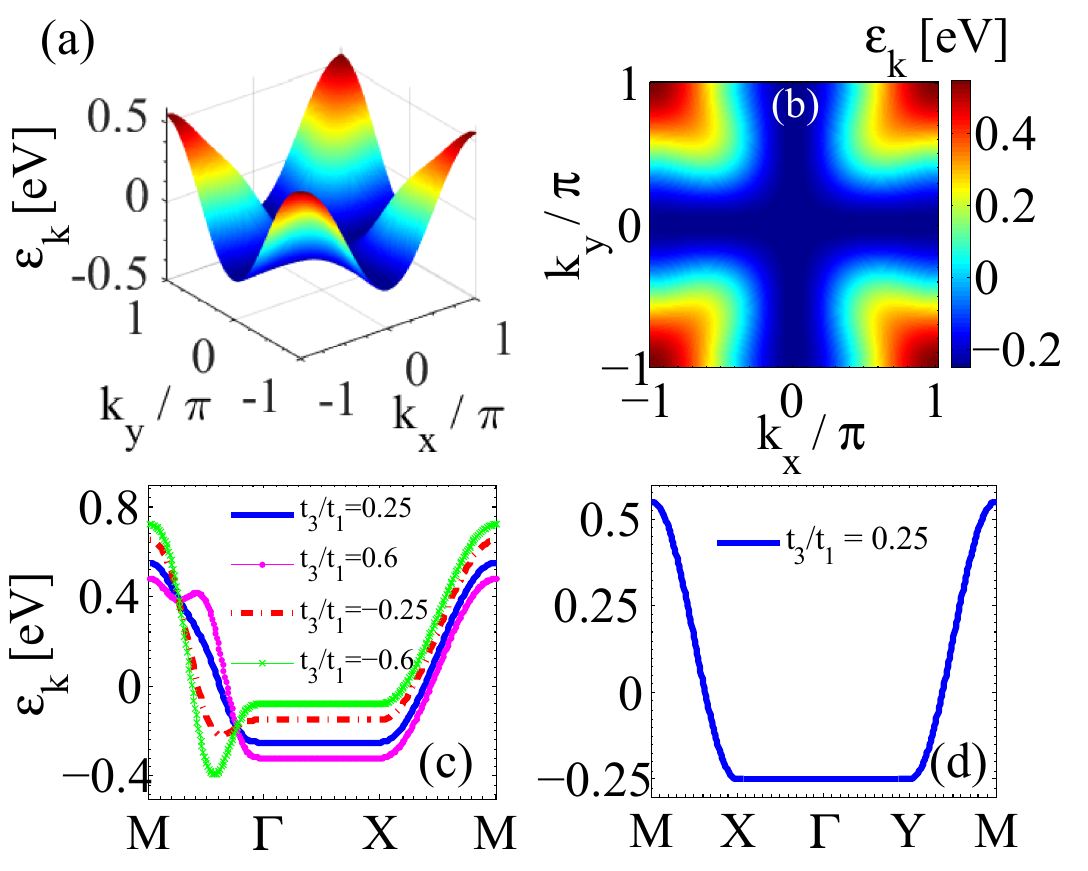}
\end{center}
\caption{\label{Fig:1}Top panel: One-electron band dispersion for a square lattice up to 5th nearest-neighbor with the hopping parameters $t_1 = 0.1$ eV, $t_2 =-0.5 t_1=-0.05$ eV, $t_3=0.25 t_1=0.025$ eV, $t_4=0$, and $t_5=-0.5 t_3=0.0125$ eV. Blue region represents constant energy (flat region). Bottom panel: The band dispersion along (c) the high symmetry line $\text{M}-\Gamma-\text{X}-\text{M}$ in the Brillouin zone for different values of the hopping parameter $t_3$ and (d) the $\text{M}-\text{X}-\Gamma-\text{Y}-\text{M}$ line, when $t_3/t_1 = 0.25$.}
\end{figure}
We consider the Hubbard model on a square lattice,
\begin{equation}\label{Hubbard_model}
\hat{H}=\sum_{\bm k, \sigma}\varepsilon_{\bm k} \hat{c}_{\bm k,\sigma}^{\dag} \hat{c}_{\bm k,\sigma}+U \sum_i \hat{n}_{i\uparrow} \hat{n}_{i\downarrow}-\mu_0 \sum_{i,\sigma} \hat{n}_{i\sigma},
\end{equation}
where the operator $\hat{c}_{\bm k,\sigma}^{\dag}$ ($\hat{c}_{\bm k,\sigma}$) creates (annihilates) an electron with momentum $\bm k=(k_x,k_y)$ and spin $\sigma$, $\varepsilon_{\bm k}$ is the non-interacting band dispersion, $\hat{n}_{i\sigma} = \hat{c}_{i\sigma}^{\dag} \hat{c}_{i\sigma}$ represents the number of electrons of spin $\sigma$ at site $i$, $U$ is the strength of the on-site Hubbard interaction, and $\mu_0$ is the chemical potential.

The non-interacting band dispersion for the square lattice up to fifth-nearest-neighbour can be written as
\begin{eqnarray}
\label{nonintdispersion}
\varepsilon_{\bm k}&=&-2t_1[\cos (k_xa)+\cos (k_ya)]-4 t_2 \cos (k_xa) \cos (k_ya)\nonumber\\
&-&2 t_3 [\cos (2k_xa)+\cos (2k_ya)]-4 t_4 [\cos (2k_xa) \cos (k_ya)\nonumber\\
&+&\cos (k_xa) \cos (2k_ya)]-4 t_5 \cos (2k_xa) \cos (2k_ya),
\end{eqnarray}
where $t_j\ (j=1-5)$ is the jth-nearest-neighbor hopping parameter and $a$ is the lattice constant.

Fig. \ref{Fig:1}(a-b) shows the band structure of the square lattice in its non-interacting limit. Importantly, it is seen that we can flatten the dispersion along $\Gamma-\mathrm{X}$ and $\Gamma-\mathrm{Y}$ directions, by setting the hopping parameters $t_2 =-0.5 t_1$, $t_3=0.25 t_1$, $t_4=0$, and $t_5=-0.5 t_3$. Setting the nearest-neighbor hopping parameter $t_1=0.1$ eV, we can express the parameters as $t_2 =-0.05$ eV, $t_3=0.025$ eV, and $t_5=-0.0125$ eV. We have further shown that tuning the sign or magnitude of the hopping parameter $t_3$ leads to the appearance of a deep or a peak for dispersive region along $\mathrm{M}-\Gamma$ direction, whereas the flat region is not changed [see Fig.\ref{Fig:1}(c)]. Therefore, by controlling the hopping parameters, we will have a partially flat-band (PFB) model on a square lattice. We should mention that the field of cold atoms on artificial lattices offers a platform allowing the direct tuning of the physical parameters of the model Hamiltonian.
The proposed PFB model can be realized in $\tau$-type organic salt family~\cite{organic1,organic2,PFB0}, D$_2$A$_1$A$_y$, based on D (=P-S, S-DMEDT-TTF or EDO-S, S-DMEDT-TTF) in combination with anions A (=AuBr$_2$, I$_3$, or IBr$_2$), which are two-dimensional metals in the $\tau$-crystal form and presents a flat-bottomed band structure, similar to our model.

\begin{figure}[]
\begin{center}
\includegraphics[width=3.3in]{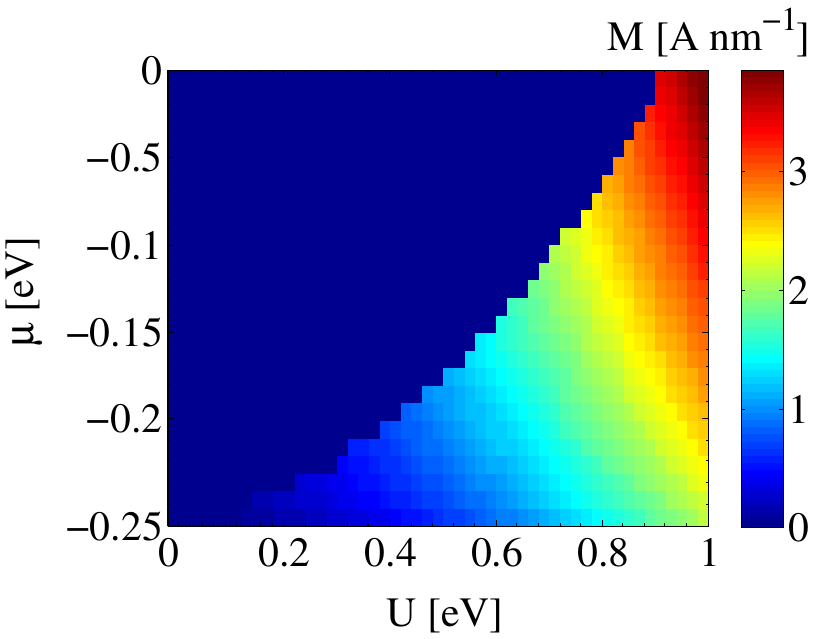}
\end{center}
\caption{\label{Fig:2}The phase diagram of the partially flat-band (PFB) system on a square lattice with $t_1 = 0.1$ eV, $t_2 =-0.5 t_1=-0.05$ eV, $t_3=0.25 t_1=0.025$ eV, $t_4=0$, and $t_5=-0.5 t_3=-0.0125$ eV, at zero temperature.}
\end{figure}
To explore magnetism in our model, we use a path integral approach to itinerant magnetism by making use of a Hubbard-Stratonovich method~\cite{Coleman}. The path integral expression for the partition function is written as $Z=\int D[c,\bar{c}] e^{-S}$, where
\begin{eqnarray}
\label{action1}
S=\int_0^{\beta} d\tau \left[\sum_{\bm{k},\sigma}\bar{c}_{\bm{k}\sigma}\left(\partial_{\tau}+E_{\bm k}\right) c_{\bm{k}\sigma}-\frac{I}{2}\sum_j(\bm{\sigma}_j)^2\right]
\end{eqnarray}
is the action, $\bar{c}_{\bm{k}\sigma}$ and $c_{\bm{k}\sigma}$ are anticommuting Grassman numbers, $E_{\bm k}=\varepsilon_{\bm k}-\mu$ and the coupling constant is introduced as $I=U/3$. Note that we have rewritten the Hubbard interaction in terms of the spin operators $U \hat{n}_{i\uparrow} \hat{n}_{i\downarrow}=-U(\bm{\sigma}_j)^2/6+U(\hat{n}_{i\uparrow}+ \hat{n}_{i\downarrow})/2$, in which the second term can be absorbed into the redefinition of the chemical potential by writing $\mu = \mu_0 - U/2$. Using the Hubbard-Stratonovich transformation by adding a white-noise field $\bm{m}_j$ into the action \eqref{action1}, $-I\sum_j(\bm{\sigma}_j)^2/2\rightarrow -I\sum_j(\bm{\sigma}_j)^2/2+\sum_j \bm{m}_j^2/2I$, and shifting $\bm{m}_j=\bm{M}_j-I\bm{\sigma}_j$ ($\bm{M}_j$ is a fluctuating field), the transformed partition function is obtained as $Z=\int D[\bm M] e^{-S_E[\bm M]}$ with ~\cite{Coleman}
\begin{eqnarray}\nonumber
&&S_E[\bm M]=\\ \nonumber
&&-\ln\left\{\int D[c,\bar{c}]\exp\left(-\int_0^\beta d\tau \left[\bar{c} (\partial_{\tau}+h_E[\bm{M}]) c
+\sum_j\frac{\bm{M}_j^2}{2I}\right]\right)\right\}.\\
\label{action2}
\end{eqnarray}

Here, $S_E[\bm M]$ is the effective action associated with a particular space-time configuration of the magnetization $\bm{M}_j(\tau)$ and $[h_E]_{\bm{k'},\bm{k}}=E_{\bm k}\delta_{\bm{k'},\bm k}-\bm{M}_{\bm {k'}-\bm k}(\tau)\cdot\bm{\sigma}$ describes the effective Hamiltonian for the electrons moving in the (time-dependent) magnetization field. Therefore, the interacting problem can be transformed, by the Hubbard-Stratonovich method, into a problem of free electrons moving in a fluctuating effective field.
\begin{figure}[]
\begin{center}
\includegraphics[width=3.3in]{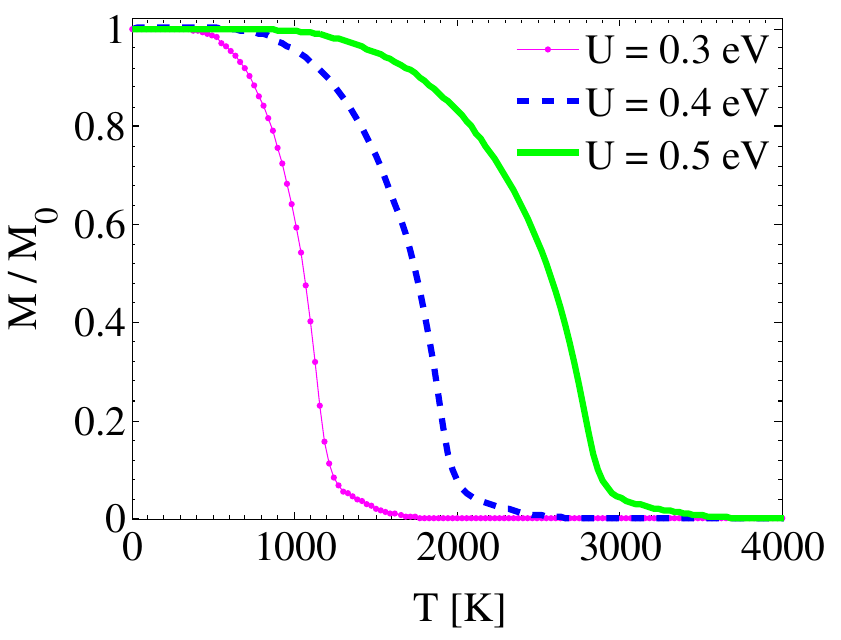}
\end{center}
\caption{\label{Fig:3}The behavior of the normalized magnetization of the proposed PFB model in terms of the temperature $T$ for three values of the Hubbard interaction, when the chemical potential resides inside the flat portion of the band structure. The zero-temperature magnetization is defined by $M_0=0.255, 0.419$, and $0.621$ A/nm, respectively for $U=0.3,0.4$ and $0.5$ eV, and the temperature is in units of Kelvin (K).}
\end{figure}
Making a saddle-point approximation, approximating the partition function by its value at the saddle point $\bm M = \bm M^{(0)}$, the effective action is directly related to the mean-field partition function $e^{-S_E[\bm{M}^{(0)}]}=Tr[e^{-\beta {{\hat{H}}}_{MF}}]$ with ${\hat{H}}_{MF}=c^{\dag}h_E[\bm{M}^{(0)}]c+\sum_j (\bm{M}_j^{(0)})^2/2I$.

To study the phase diagram of the proposed partially flat-band model, we suppose a uniform magnetization along the $z$-axis, $\bm{M}_j^{(0)}= M \hat{\bm z}$. Carrying out the Gaussian integral over the Fermi fields [Eq. (\ref{action2})], we can write the effective action in a Fourier space as
\begin{eqnarray}
S_E[M]=-\sum_{\bm k,\sigma} \ln\left[1+e^{-\beta(E_{\bm k}-\sigma M)}\right]+N_s\beta\frac{M^2}{2I}.
\end{eqnarray}

The stationary point of the action can be achieved by differentiating the free energy per unit volume $F_E$ ($F_E = S_E/\beta N_s$) with respect to $M$ ($-\partial F_E/\partial M = 0$), which expresses the mean-field condition
\begin{equation}
\label{magnetization}
M=I\sum_{\bm{k},\sigma=\pm 1}\sigma f(E_{\bm k}-\sigma M),
\end{equation}
with $f(x)=(1+e^{\beta x})^{-1}$, the Fermi-Dirac distribution function. Solving the above equation self-consistently, we can calculate the magnetization of the partially flat-band system in terms of the strength of the Hubbard interaction $U$ and the chemical potential $\mu$. Before presenting our results, we should emphasize that the interaction-induced mean-field magnetization $M$ will be scaled in terms of $\mu'_0 \mu_B$ (A/m), with $\mu_B=5.788\times10^{-5}$ (eV/Tesla) the Bohr magneton and $\mu'_0=4\pi\times10^{-7}$ (Tesla/A) the magnetic permeability of vacuum. All energies are in units of electron volt (eV). The temperature T is in units of Kelvin (K). Also, the lattice constant $a$ is set to unity.

First, we determine the characteristic magnetism of the present PFB model based on the phase diagram using Eq. \eqref{magnetization}. Fig. \ref{Fig:2} shows the phase diagram against the strength of the Hubbard interaction $U$ and the chemical potential $\mu$, at zero temperature. According to the Stoner theory of itinerant magnetism in metals, strong interactions drive a metal to become unstable towards the development of a spontaneous spin polarization. Surprisingly, we demonstrate that the suppression of the electronic kinetic energy in the flat portion of the band dispersion (with $\mu=-0.25$ eV) leads to the appearance of ferromagnetism even at very weak interaction regime. However, reducing the magnitude of the chemical potential by going away from the flat region (dispersive region) makes the non-magnetic (metal or Mott insulator)-magnetic transition to be occurred at higher values of $U$.

In order to better grasp the magnetic phase diagram, we have presented the colormap of the magnetization in terms of the Hubbard interaction and the band filling $n_0$ in Fig. \ref{Fig:A3} [see Appendix \ref{sec:appendix A}]. It can be seen that the white dotted-line, which corresponds to the line $\mu_0=U/2-0.25$ [$\mu=-0.25$ eV], determines the $U$-dependent mean-filed magnetization for the band filling $n_0<1.63$ (per unit cell) when the non-interacting chemical potential $\mu_0$ resides inside the flat portion of the band structure.

We further present the temperature dependence of the magnetization $M$ (normalized to the zero-temperature magnetization $M_0$) in the weakly interacting regime, when the zero-temperature chemical potential is located inside the flat portion of the spectrum [see Fig. \ref{Fig:3}]. This is computed by simultaneous solution of the mean-field magnetization equation [Eq. (\ref{magnetization})] and the temperature-dependent equation for the chemical potential, obtained by imposing the constant-electron density condition $n(T)=n(T=0)$ with $n=\sum_{\bm{k},\sigma=\pm 1}f(E_{\bm k}-\sigma M)$. Interestingly, it is found that the magnetic transition temperature $T_C$ is about $2000^\circ$K in the weakly interacting regime ($U=0.3$ eV), which is considerably larger that that of other materials. Moreover, enhancing the strength of the Hubbard interaction leads to a significant amplification of the transition temperature $T_C$ such that it reaches $4000^\circ$K for $U=0.5$ eV.

In addition, we have evaluated the phase diagram for another set of hopping parameters ($t_1=0.1$ eV, $t_2=-0.5 t_1=-0.05$ eV, and $t_{j>2}=0$) resulting in partially flat-band $t-t'$ model, as well as the dispersive regular band model in Appendix \ref{sec:appendix B}. We have shown that the larger portion of the non-dispersive flat region in the $t-t'$ model (in analogy with our model) makes the magnetic state to be appeared for a wider window of the chemical potential $\mu$ and the interaction $U$. Comparing the results with that of a dispersive regular band model confirms that the presence of flat bottom below the dispersive part of the band is responsible for the appearance of Stoner magnetism at very weak interaction regime.

\begin{figure}[]
\begin{center}
\includegraphics[width=3.4in]{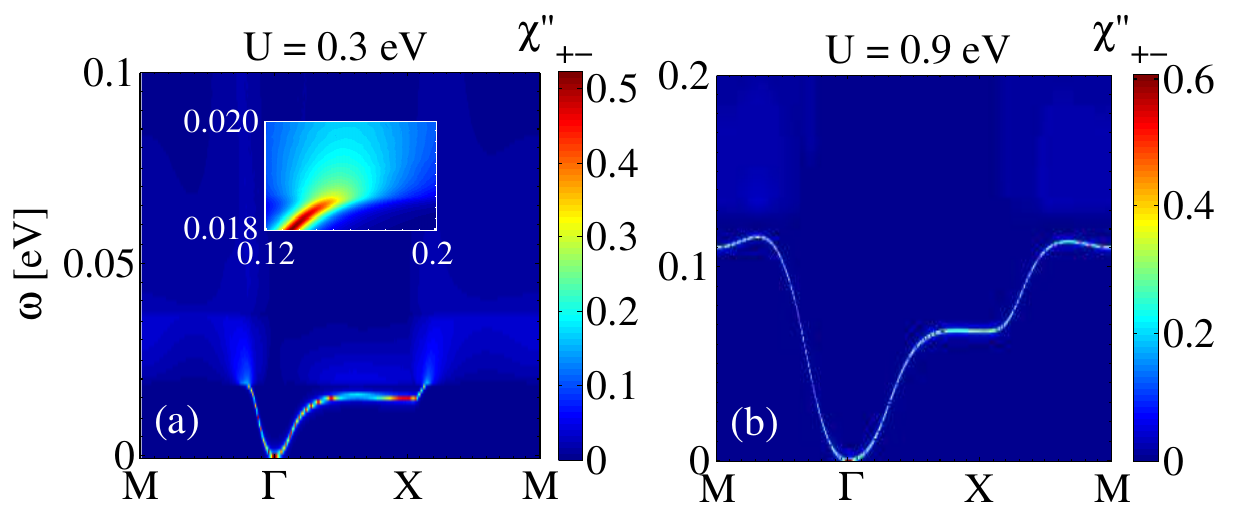}
\end{center}
\caption{\label{Fig:4}Density plot of the dynamical transverse spin susceptibility along the M-$\Gamma$-X-M direction for two values of the Hubbard interaction $U=0.3$ eV, with the magnetization $M=0.231$ A/nm (a) and $U=0.9$ eV, with $M=1.748$ A/nm (b) at zero temperature, when the chemical potential $\mu=-0.25$ eV. Inset of (a) shows the zoomed in view of the dynamical spin susceptibility near the Stoner continuum.}
\end{figure}

\section{magnetic collective modes}\label{collective}
\begin{figure*}[]
\begin{center}
\includegraphics[width=6in]{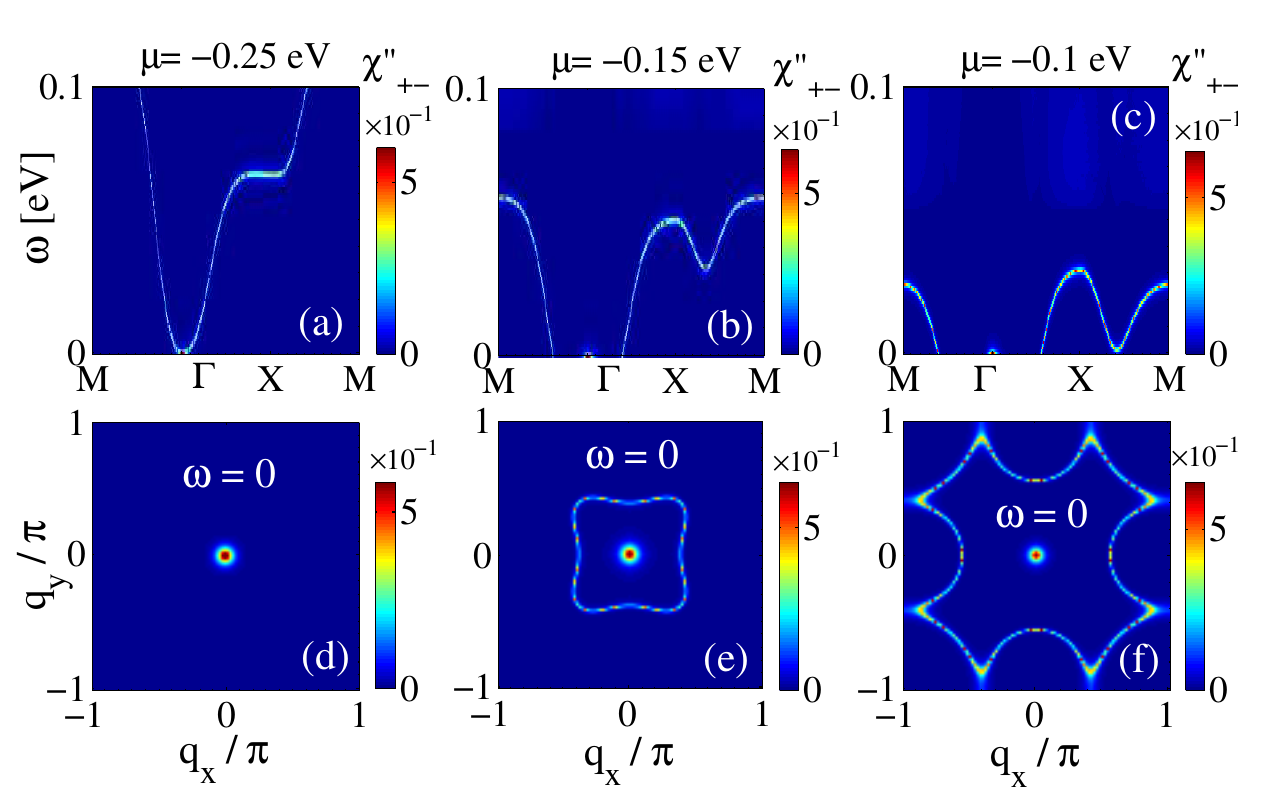}
\end{center}
\caption{\label{Fig:5}Top (bottom) panel: The dynamical (static) transverse spin susceptibility along the $\text{M}-\Gamma-\text{X}-\text{M}$ direction for three values of the chemical potential $\mu=-0.25, -0.15$ and $-0.1$ eV at zero temperature, when $U=0.9$ eV.}
\end{figure*}

Now, we go beyond the mean-field theory to evaluate the quantum fluctuations in magnetization by expanding the magnetization in fluctuations around the saddle point
\begin{equation}
\bm M_j(\tau) = \bm M^{(0)} + \delta \bm M_j(\tau),
\end{equation}
or in Fourier space $\bm M_q = \bm M^{(0)}\delta_{q=0} + \delta M_q$ with $q\equiv(\bm q, i\nu_n)$. Expanding the effective action up to second order in the fluctuations and using $M_{k-k'}= M\delta_{k-k'}+\delta M_{k-k'}$, we obtain
\begin{eqnarray}
\hspace{-5mm}F_E[\bm M] &=& -\frac{1}{N_s \beta} \text{Tr}\ \ln\left[-G(k)^{-1}\delta_{k,k'}-\delta \bm M_{k-k'}\cdot\bm{\sigma}\right]\nonumber\\
&+&\sum_q \frac{|M\bm{\hat{z}}\delta q+\delta \bm M_q|^2}{2 I},
\end{eqnarray}
where $G(k)=(i\omega_n-E_{\bm k}-\sigma_z M)^{-1}$ is the renormalized propagator, and $\omega_n$ is the Matsubara frequency. Therefore, the Gaussian action for the magnetization fluctuations takes the form
\begin{equation}
\Delta F_G[\bm M]=\frac{1}{2}\sum_q \delta M_{-q}^a\left[\frac{\delta_{ab}}{I}-\chi_{ab}^{(0)}(q)\right] \delta M_{-q}^b ,
\end{equation}
with
\begin{equation}
\label{baresusceptibility}
\chi_{ab}^{(0)}(q)=-\frac{1}{N_s\beta}\sum_k Tr [\sigma_a G(k + q) \sigma_b G(k)]
\end{equation}
the bare susceptibility of the polarized metal. Using $\sigma_{\pm}= (\sigma_x \pm i\sigma_y)/2,$ and $M_q^{\pm}= M_q^x\pm i M_q^y$ for diagonalizing the magnetic fluctuations and the distribution function $p[M_q] \propto e^{-\beta N_s \Delta F_G [\bm M]}$ for the Gaussian magnetic fluctuations about the Stoner mean-field theory for an itinerant ferromagnet, the longitudinal and transverse fluctuations  can be written as
\begin{equation}
\langle\delta M_q^z\ \delta M_{-q}^z\rangle=\frac{1}{I^{-1}-\chi_{zz}^{(0)}(q)},
\end{equation}
and 
\begin{equation}
\langle\delta M_q^+\ \delta M_{q}^-\rangle=\frac{1}{{(2I)}^{-1}-\chi_{+-}^{(0)}(q)},
\end{equation}
respectively.

Recalling the Hubbard-Stratonovich transformation [$\bm{M}_j(\tau)=\bm{m}_j(\tau)+I\bm{\sigma}_j(\tau)$], the transverse random phase approximation (RPA) spin fluctuations of an itinerant ferromagnet takes the form~\cite{Coleman}
\begin{equation}
\label{susceptibility}
\chi_{+-}(q)=\frac{\chi_{+-}^{(0)}(q)}{1 - 2 I \chi_{+-}^{(0)}(q)}.
\end{equation}
Finally, employing Eqs. (\ref{baresusceptibility}) and (\ref{susceptibility}), we can evaluate the transverse spin fluctuations of the magnetized partially flat-band model with the transverse bare susceptibility
\begin{eqnarray}
\chi_{+-}^{(0)}(q)&=&-\frac{1}{N_s\beta}\sum_{\bm k,i\omega_n} Tr [\sigma_+ G(k + q) \sigma_- G(k)]\nonumber\\
&=&-\frac{1}{N_s\beta}\sum_{\bm k,i\omega_n}G_{\uparrow}(k + q)G_{\downarrow}(k),\\
\chi_{+-}^{(0)}(\bm{q},\nu)&=&\frac{a^2}{(2\pi)^2}\sum_{\sigma=\pm 1} \int d^2k f(E_{\bm k}-\sigma M)\nonumber\\
&\times&\frac{1}{E_{\bm k+\bm q}-E_{\bm k}-\sigma (\nu-2 M)}.
\end{eqnarray}

Keeping the phase diagram in mind, we know discuss the energy spectrum of quantum magnetic fluctuations in the proposed PFB model. Fig. \ref{Fig:4}(a) shows the  density plot of the dynamical transverse spin susceptibility $\chi_{+-}''(\bm q,\omega)=\text{Im}\ \chi_{+-}(\bm q,\omega-i\delta)$ predicted by the RPA theory along the $\text{M}-\Gamma-\text{X}-\text{M}$ direction for the Hubbard interaction $U=0.3$ eV, with the magnetization $M=0.231$ A/nm, when the chemical potential resides inside the flat region ($\mu=-0.25$ eV). The low-energy excitations are dispersive magnon modes with strong intensity around the $\Gamma$ point, exhibiting the ferromagnetic correlations. Notably, a flat magnon band appears at high energies along the $\Gamma-\text{X}$ direction, similar to that of the electronic spectrum, which itself is connected to the Stoner continuum via another dispersive part. As illustrated in Fig. \ref{Fig:4}(b), enhancing the interaction strength leads to the appearances of a flat region with smaller length at higher energies in comparison with that of $U=0.3$ eV.
\begin{figure}[]
\begin{center}
\includegraphics[width=3.5in]{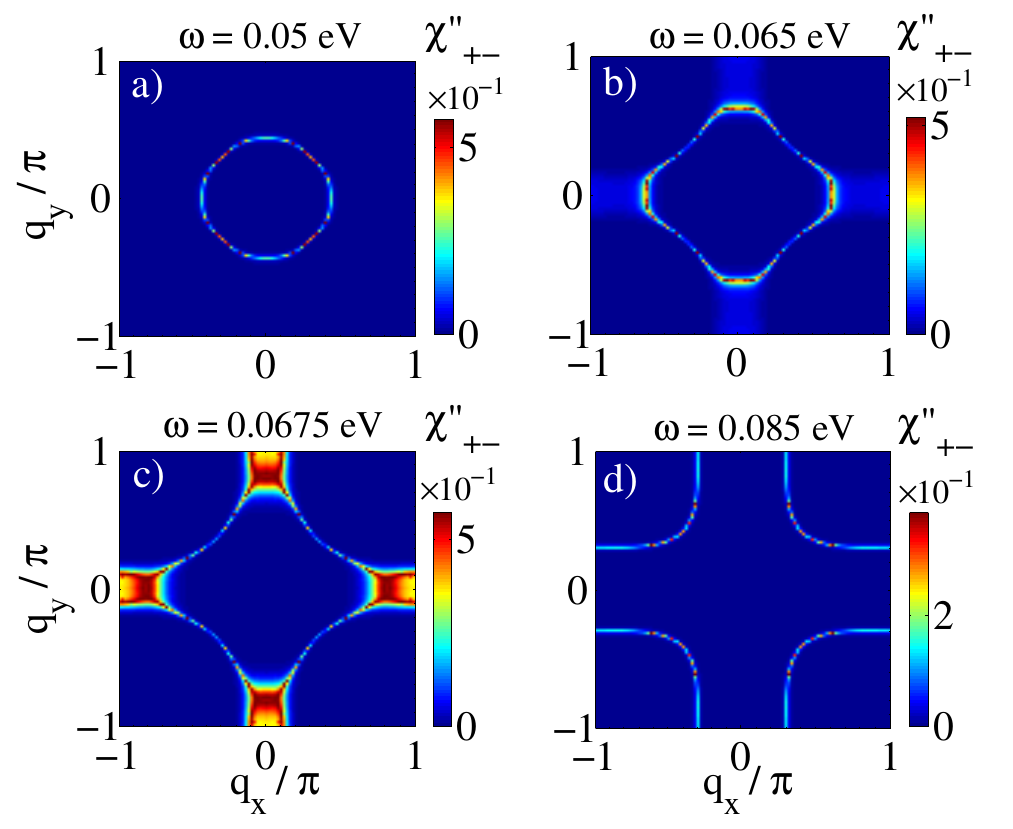}
\end{center}
\caption{\label{Fig:6}Constant-energy cuts of the dynamical transverse spin susceptibility for (a) $\omega=0.05$ eV, (b) $\omega=0.065$ eV, (c) $\omega=0.0675$ eV and (d) $\omega=0.085$ eV, when $\mu=-0.25$ eV and $U=0.9$ eV.}
\end{figure}
Most importantly, as depicted in Fig. \ref{Fig:5}, reducing the magnitude of the chemical potential at a strong interaction regime ($U=0.9$ eV) leads to an additional magnetic order (complex spin density wave) at finite wave vectors. In addition, it can be seen that the flat part of the magnetic excitations disappears when the chemical potential placed outside the flat region of the electronic spectrum.

To achieve qualitative understanding of magnetic excitations, we present the constant-energy cuts through the excitation spectra in Fig. \ref{Fig:6}. Starting at $\omega=0$, we observe hot spot at $\Gamma$ point, which is an evidence of ferromagnetic correlations [see Fig. \ref{Fig:5}(d)]. As we move up in energy, the dispersive excitations form ring of intensity around the $\Gamma$ zone which eventually forms diamond with streaks of intensity along the $\Gamma$-X and $\Gamma$-Y lines. As we continue to increase energy, the spin correlation is seen to be large over the streaks or wide plateaus (rather than the usual spot), which should come from the flattened band. These results can be probed experimentally using inelastic neutron scattering measurements. We should note that flat magnon bands are observed in insulating kagome ferromagnet Cu~\cite{kagomecu} and kagome metal TbMn$_6$Sn$_6$~\cite{kagomeTb}.

Besides, we have evaluated the magnetic excitation spectrum for the partially flat-band $t-t'$ model in Appendix \ref{sec:appendix C} and shown that, in contrast to the electronic counterpart, it has smaller flat magnon band than that of the proposed PFB model. Also, we have found (not shown) that the static transverse spin susceptibility displays stronger ferromagnetic ordering at the $\Gamma$ point in the $t-t'$ model.

\section{Thermoelectric properties of the correlated metallic phase}\label{thermoelectric}
\begin{figure*}[]
\begin{center}
\includegraphics[width=7in]{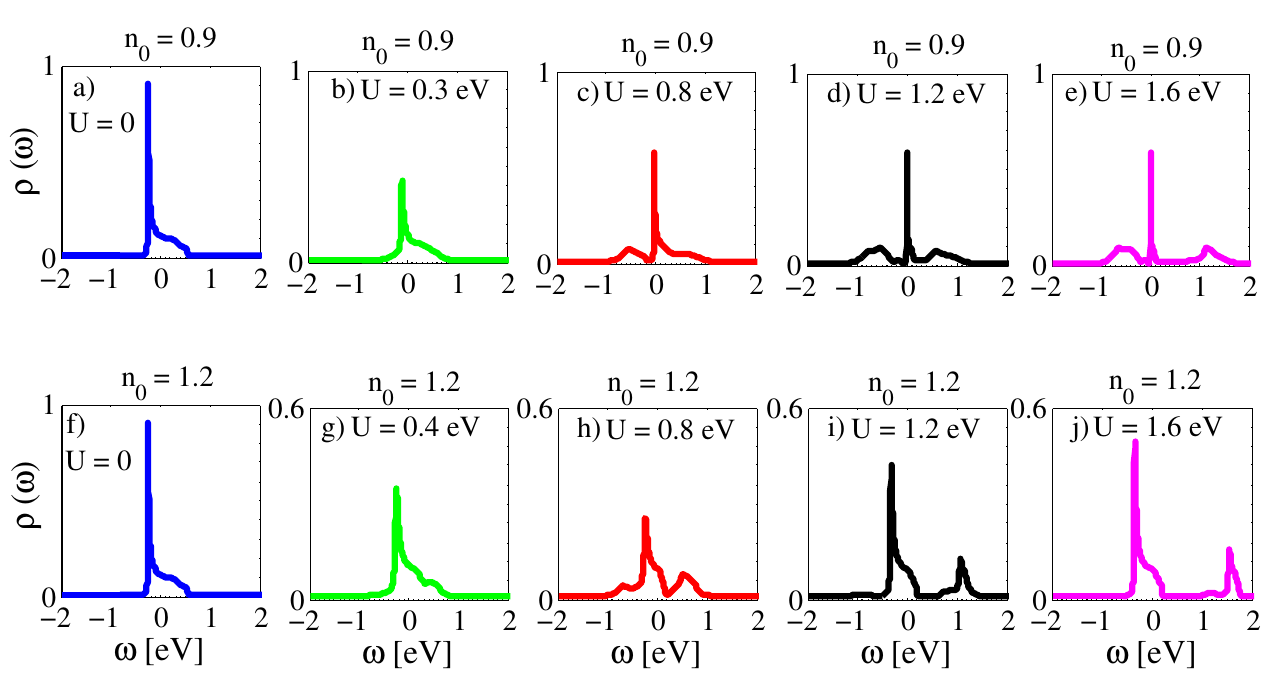}
\end{center}
\caption{\label{Fig:7} Top (bottom) panel: The behavior of the density of states for different values of the Hubbard interaction $U$, when the band filling is $n_0 = 0.9$ ($n_0 = 1.2$), the bandwidth is $W=0.8$ eV, and the temperature is set to zero.}
\end{figure*}
In this section we explore the partially flat-band model \eqref{Hubbard_model} in the absence of ferromganetic instability, assuming that the model is a correlated metal. In particular, we investigate the thermoelectric properties of the system in a wide range of interaction strengths using dynamical mean field theory (DMFT)~\cite{DMFT} with a modified version of iterative perturbation theory as a solver to calculate the self-energy~\cite{DMFT_IPT}. DMFT is a method based on mapping a many-body lattice problem (the Hubbard model) onto a many-body local problem, the so-called Anderson impurity model describing the interaction of one site (the impurity) with a bath of electronic levels through a hybridization function $\Delta$. This hybridization function is related to the non-interacting Green's function of the bath by the relation
\begin{equation}
G_{\text{bath}}(i\omega_n)=\frac{1}{i\omega_n+\mu_{\text{bath}}-\Delta(i\omega_n)},
\end{equation}
where $\mu_{\text{bath}}$ is a fictitious chemical potential of the bath, determined by imposing sum rules or by other methods. Note that the chemical potential of the bath $\mu_{\text{bath}}$ is the same as the chemical potential of the non-interacting system $\mu_0$, and therefore the corresponding band filling $n'_0$ is equal to $n_0$. The hybridization function should reproduce the lattice dynamics such that the impurity Green's function is the same as the local lattice Green's function. Therefore, the self-consistency condition connects the Green's function of the local impurity with on-site energy $\varepsilon_{imp}$, 
\begin{equation}
G_{\text{imp}}^{-1}(i\omega_n)=G_{\text{bath}}^{-1}(i\omega_n)-\mu_{\text{bath}}-\varepsilon_{\text{imp}}-\Sigma(i\omega_n),
\end{equation}
and the lattice Green's function with the non-interacting electronic dispersion of $\varepsilon_{\bm k}$ on the lattice [Eq. (\ref{nonintdispersion})] and the chemical potential $\mu'$
\begin{equation}
G'(i\omega_n)=\sum_{\bm k} \frac{1}{i\omega_n-\varepsilon_{\bm k}+\mu'-\Sigma(i\omega_n)},
\end{equation}
by imposing $G'(i\omega_n) = G_{\text{imp}}(i\omega_n)$ and $\mu' = -\varepsilon_{\text{imp}}$. Therefore, we have a closed set of equations which must be iterated until numerical convergence is reached. We note that the existence of on-site correlations on the impurity site leads to a nonzero local self-energy~\cite{self_energy}
\begin{equation}
\label{selfenergy}
\Sigma(\omega)=U n'+\frac{A \Sigma^{(2)}(\omega)}{1-B \Sigma^{(2)}(\omega)}.
\end{equation}

Here, $\Sigma^{(2)}(i\omega_n)=-U^2\int_0^\beta d\tau e^{i\omega_n\tau} G_{\text{bath}}^2(\tau) G_{\text{bath}}(-\tau)$, $A= n'(1-n')/n'_0(1-n'_0)$, $B =[(1-n')U+\mu_{\text{bath}}-\mu']/n'_0(1-n'_0)U^2$, $n'_0=-\pi^{-1}\int_{-\infty}^{\infty} f(\omega) \text{Im} G_{\text{bath}}(\omega) d\omega$, and $n'=-\pi^{-1}\int_{-\infty}^{\infty} f(\omega)\text{Im} G'(\omega) d\omega$. Also, the quantities $\mu_{bath}$ and $\mu'$ are fixed by the filling and the Friedel sum rule~\cite{Friedel} or the Luttinger theorem~\cite{Luttinger}.

Having found the interacting Green's function $G'(\omega)$ in self-consistency loop, by using the self-energy relation in Eq. (\ref{selfenergy}), we can calculate the electrical and thermal responses of the system such as the dc conductivity $\sigma$, the thermal conductivity $\kappa$, Seebeck coefficient $S$, that relates the gradients of temperature and electrical field, and the figure of merit $ZT=\sigma S^2 T/\kappa$. Within the Kubo formalism~\cite{Kubo}, they can be expressed in terms of current-current ($A_0$) and current-heat correlation  ($A_1$) functions as
\begin{equation}
\sigma=\frac{2\pi e^2}{\hbar}A_0,
\end{equation}
\begin{equation}
S=-\frac{k_B}{|e|}\frac{A_1}{A_0},
\end{equation}
\begin{equation}
\kappa=\frac{2\pi k_B^2}{\hbar}T \left(A_2-\frac{A_1^2}{A_0}\right)
\end{equation}
with $A_n=\int_{\infty}^{\infty} d\omega \beta^n \omega^n (-{\partial f}/{\partial \omega}) \Xi(\omega)$ the correlation function, $\Xi(\omega)=\int_{-\infty}^{\infty} d\varepsilon \rho(\omega,\varepsilon)^2 \Phi(\varepsilon)$ the transport kernel, $\Phi(\varepsilon)=\Sigma_{\bm k}(\partial\varepsilon_{\bm k}/\partial k_x)^2 \delta(\varepsilon_{\bm k}-\varepsilon)$ the transport function and $\rho(\omega,\varepsilon)=-\pi^{-1} \text{Im} [\omega-\varepsilon+\mu'-\Sigma(\omega)]^{-1}$ the density of states.

\subsection{Density of states}

Now, we proceed to investigate the behavior of the density of states $\rho(\omega)$ in the proposed PFB model. The top and bottom panels of Fig. \ref{Fig:7}, respectively, presents the zero-temperature density of states for different values of the on-site Hubbard interaction, when the band fillings are $n_0=0.9$ and $1.2$, and the bandwidth is $W=0.8$ eV. Obviously, enhancement of the Hubbard interaction drives the system from a normal metallic phase to a correlated metallic state, when $U$ exceeds the bandwidth $W$, otherwise a Mott insulator phase at half-filling. The sharp peaks at $U=0$ in panels Fig.~\ref{Fig:7}(a) and (f) originates from the flat portion of the bands with dirverging density of states. By increasing the Hubbard interaction two features appear at higher energies away from $\omega=0$, which manifest Mott bands. Away from the half-filling, yet close to the flat region, e.g., in the case of $n_0=0.9$, the density of states is sharply peaked around $\omega=0$, manifesting the effect of the flatness of the electronic spectrum. However, when the doing level is far away from the flat region, e.g., the case of $n_0=1.2$, there is no evidence of flat band induced large density of states as seen in Fig.~\ref{Fig:7}(g-j).

\subsection{Seebeck coefficient and figure of merit}
\begin{figure}[t]
\begin{center}
\includegraphics[width=3.5in]{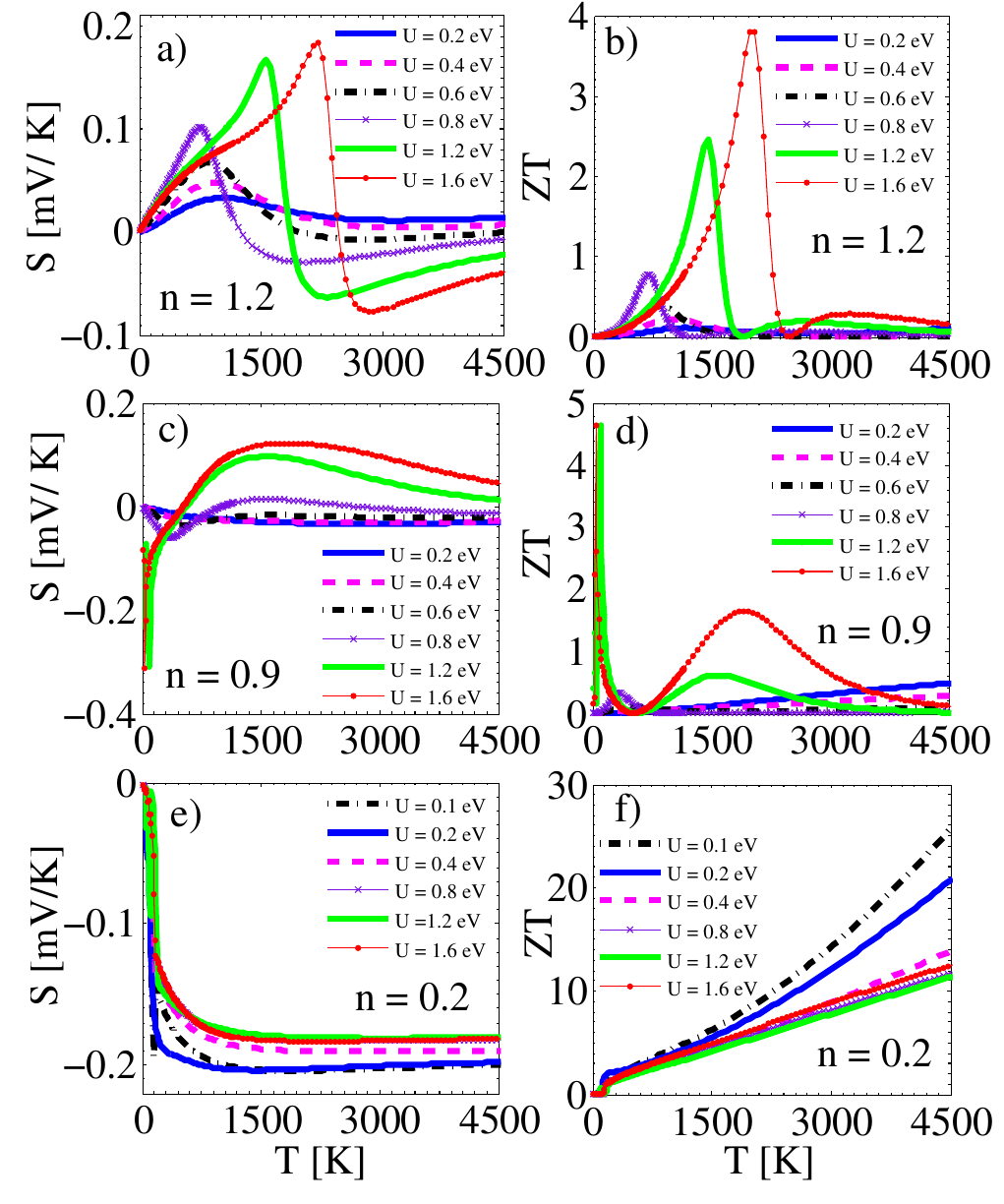}
\end{center}
\caption{\label{Fig:9}Left (right) panel: Seebeck coefficient (figure of merit) in terms of the temperature for three values of band filling $n_0=1.2, 0.9$ and $0.2$, when the bandwidth is set to $0.8$ eV.}
\end{figure}
Finally, we evaluate the effect of electronic correlations and specially the flatness of electronic spectrum on the thermoelectric properties of the proposed model by focusing on the seebeck coefficient $S$ and figure of merit $\text{ZT}$. Note that we only consider the electronic contributions to the transport coefficients and ignore the phonons. The left and right panels of Fig. \ref{Fig:9}, respectively, show the variation of the seebeck coefficient (in units of $mV/K$) and figure of merit with the temperature for three values of band filling $n_0=1.2,0.9$ and $0.2$ at different values of the Hubbard interaction. It is seen that the seebeck coefficient increases over a wide range of temperatures by enhancing the correlations. In the case of electron doping ($n_0=1.2$), both the seebeck coefficient and the figure of merit present a maximum, which shifts to higher temperatures by enhancing the correlations. Interestingly, the seebeck coefficient changes sign (from positive to negative values) for $U\geq 0.6$ eV, which is not present in the non-interacting system. Moreover, significant amplification of the figure of merit is obvious for $U>W$ in which the system is in the correlated metallic phase.

However, in the case of hole doping (with $n_0 = 0.9$), the sign change of the seebeck coefficient (from negative to positive values) appears at stronger interaction regime with $U\simeq W$. In the case of small band filling $n_0=0.2$, in contrast with the above results, the seebeck coefficient and the figure of merit reduce with $U$, there is no change of sign for the seebeck coefficient, and it saturates with increasing the temperatures. Remarkably, a significant enhancement of the figure of merit (specially in the vicinity of flat band with $U=0.1$ eV) at both of low and high temperatures can be occurred by decreasing the band filling $n_0$. Therefore, the large seebeck coefficient and the figure of merit (specially in low band filling) is the advantage of the proposed PFB model over the symmetric lattices such as square and cubic lattices~\cite{DMFT_IPT}.

\section{Conclusions}\label{conclusion}
Motivated by recent progresses in discovering materials with flat band dispersion near the Fermi energy and its profound effects on emerging exotic phases, in this paper we introduced a one-band model on a square lattice with partially flat electronic dispersion. First, we have investigated the magnetic characteristics of the proposed partially flat-band (PFB) system using the Hubbard model. Thanks to the quenched kinetic energy, the PFB can host correlated electronic states and displays remarkable strongly interacting phases of matter. Importantly, we demonstrated that the proposed model has a strong tendency towards Stoner ferromagnetism even at very weak interaction regime, with significantly enhanced transition temperature. Remarkably, we have shown the appearance of a flat magnon band connected to a dispersive part, similar to that of the electronic spectrum. The length of the flat magnon band strongly depends on the interaction strength and reduces with increasing the interaction. However, tuning the magnitude of the chemical potential (or band filling) at strong interaction regime leads to the appearance of an additional magnetic order (complex spin density wave) at finite wave vectors.

We then concentrated on the correlated metallic phase in the absence of magnetism and examined the effects of electronic correlations and the flatness of the electronic spectrum on the thermoelectric properties of the proposed PFB model by evaluating the seebeck coefficient and the figure of merit. In the absence of ferromagnetic instability, the Hubbard interaction is the driving force behind the formation of Mott bands. Employing dynamical mean-field theory, we calculated the electronic and transport properties of the model. For a hole dopped model away from the half-filling, the flat portion of the band is significantly pronounced near the Fermi level, which is signified as a sharp pleak in the density of states. The Hubbard bands are formed at energies away from the Fermi level when the Hubbard interaction $U$ exceeds the bandwidth $W$, depending on the value of the band filling. For electron dopping, where the filling is greater than unity, the electron states associated with the flat band are diminished at the Fermi level. Therefore, the effects of the flat band region on the emerging correlated states are much more pronounced for a hole-dopped system. In addition, the model exhibits a nonmonotonic temperature dependence for the seebeck coefficient such that it saturates at high temperatures where the incoherent regimes set in. Moreover, the correlation-induced sign change of the seebeck coefficient occurs for $U<W$ in the case of electron doping and for $U>W$ in the case of hole doping. Most importantly, the proposed model exhibits high figure of merit in addition to the seebeck coefficient, revealing the potential of the PFB systems for thermoelectric applications. Overall, our results show that even weak interactions are sufficient to encounter evidence of strong correlation physics.

\appendix

\section {\label{sec:appendix A} Phase diagram of the proposed model versus the band filling}
\begin{figure}[]
\begin{center}
\includegraphics[width=3.2in]{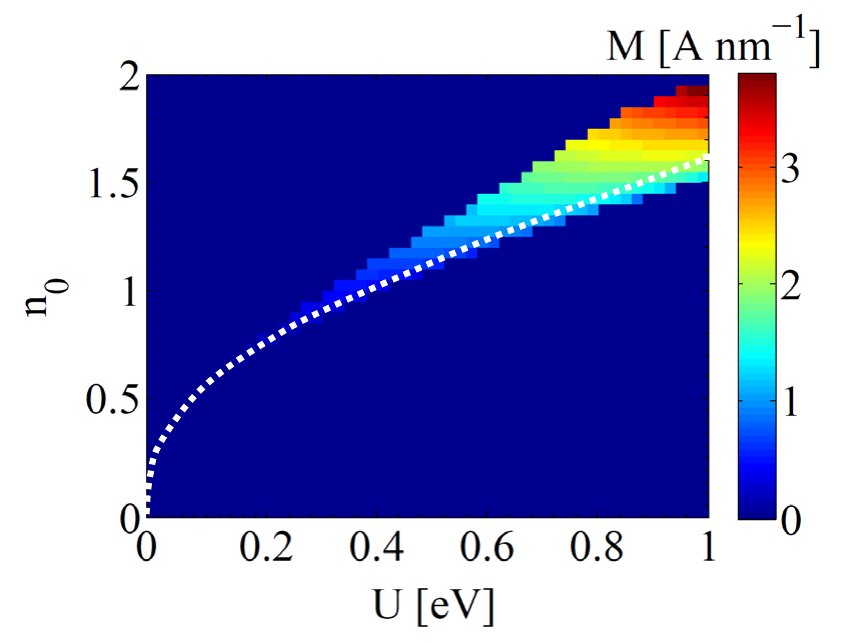}
\end{center}
\caption{\label{Fig:A3} The phase diagram of the proposed PFB system on a square lattice versus the Hubbard interaction $U$ and the band filling $n_0$ at zero temperature, when $t_1 = 0.1$ eV, $t_2 =-0.5 t_1$, $t_3=0.25 t_1$, $t_4=0$, and $t_5=-0.5 t_3$.}
\end{figure}
To express the magnetic phase transition in terms of the band filling $n_0=2\sum_{\bm{k}}f(\varepsilon_{\bm k}-\mu_0)$ at zero temperature, we have illustrated the phase diagram versus $n_0$ for a wide range of the Hubbard interaction in Fig. \ref{Fig:A3}. The white dotted-line, which corresponds to the line $\mu_0=U/2-0.25$ [$\mu=\mu_0-U/2$ with $\mu=-0.25$ eV], approves that the partially flat-band system has a tendency towards ferromagnetism even at weakly interacting regime. This occurs for the band filling $n_0<1.63$ (per unit cell), when the non-interacting chemical potential $\mu_0$ lies in a range $-0.25 eV\leq\mu_0\leq 0.25 eV$.

\section {\label{sec:appendix B} Phase diagram of the partially flat-band $t-t'$ model}

\begin{figure}[t]
\begin{center}
\includegraphics[width=3.4in]{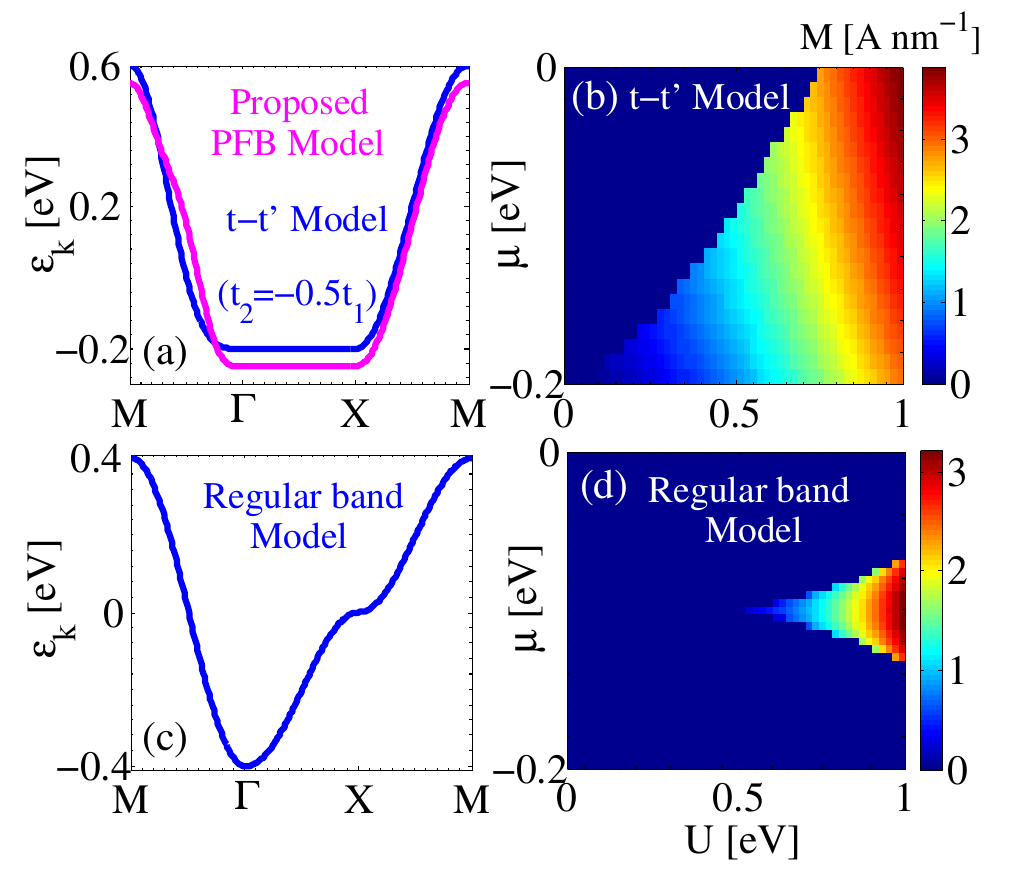}
\end{center}
\caption{\label{Fig:A1}Top panel: (a) The band structure and (b) phase diagram of the partially flat-band $t-t'$ model on a square lattice with $t_1 = 0.1$ eV, $t_2 =-0.5 t_1 = -0.05$ eV, and $t_{j>2}=0$ at zero temperature. Bottom panel: (c) The band structure and (d) phase diagram of the dispersive regular band model on a square lattice with $t_1 = 0.1$ eV, and $t_{j>1}=0$ at zero temperature.}
\end{figure}

We evaluate the colormap of magnetization for varying the Hubbard interaction $U$ and the chemical potential $\mu$ at zero temperature in the framework of mean-field theory, for another partially flat-band model as well as a dispersive model on a square lattice. The first model is the $t-t'$ model with hopping parameters $t_1 = 0.1$ eV, $t_2 = -0.5 t_1= -0.05$ eV, and $t_{j>2}=0$, which has a partially flat-band structure (the flat region is placed at $\varepsilon_{\bm k}=-0.2$ eV). The second one is a dispersive regular band model with hopping parameters $t_1 = 0.1$ eV, and $t_{j>1}=0$. If we look at the band structures in Figs. \ref{Fig:A1}(a) and \ref{Fig:A1}(c), we find that the $t-t'$ model has a larger flat region in comparison with that of our proposed PFB square model with $t_1 = 0.1$ eV, $t_2 =-0.5 t_1$, $t_3=0.25 t_1$, $t_4=0$, and $t_5=-0.5 t_3$, while the regular band model has no flat region. Therefore, comparing the phase diagrams demonstrate that, the $t-t'$ model becomes ferromagnet at smaller values of $U$ than that of our proposed structure, with enhanced magnetization [see Fig. \ref{Fig:A1} (b)]. While the absence of flat region in regular band model makes it to be at non-magnetic phase (metal or Mott Insulator) in the case of weak interacting regime [see Fig. \ref{Fig:A1} (d)].

\section {\label{sec:appendix C} Transverse spin susceptibility of the partially flat-band $t-t'$ model}

\begin{figure}[]
\begin{center}
\includegraphics[width=3.4in]{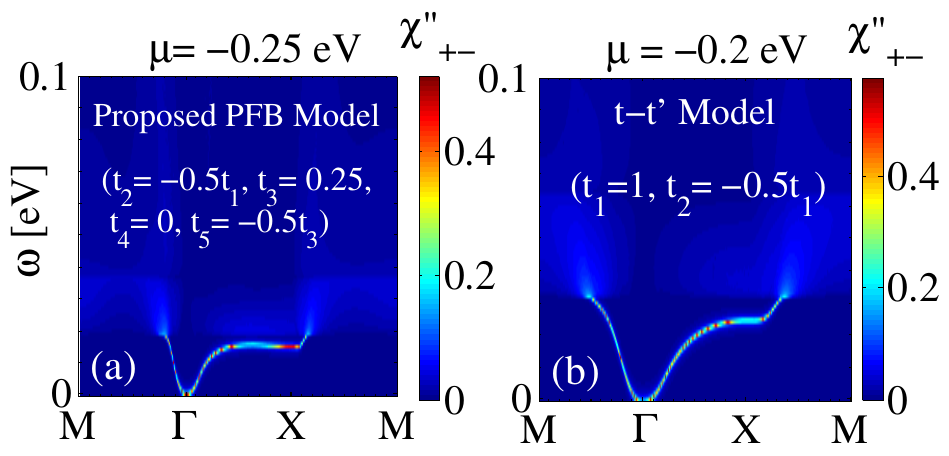}
\end{center}
\caption{\label{Fig:A2}The dynamical transverse spin susceptibility for (a) the proposed PFB model on a square lattice with $t_1 = 0.1$ eV, $t_2 =-0.5 t_1$, $t_3=0.25$, $t_4=0$, $t_5=-0.5 t_3$, and $\mu= -0.25$ eV, and (b) the $t-t'$ model on a square lattice with $t_1 = 0.1$ eV, $t_2 =-0.5 t_1$, $t_{j>2}=0$, and $\mu= -0.2$ eV at zero temperature, when $U=0.3$ eV.}
\end{figure}

We compare the magnetic excitation spectrum of the partially flat-band $t-t'$ model (right panel of Fig. \ref{Fig:A2}) with that of the proposed PFB model on the square lattice (left panel of Fig. \ref{Fig:A2}), when the chemical potential located at the flat region of the electronic spectrum and $U=0.3$ eV. We find that the behavior of the magnonic spectrum is similar, except that the flat magnon band appears at higher energies and a has smaller length than that of the proposed PFB model. This is in contrast to the results of the electronic spectrum [see Fig. \ref{Fig:A1}(a)], in which the flat portion of the $t-t'$ model has larger length than the proposed model. Also, we have found (not shown) that the transverse spin susceptibility shows stronger magnetic correlations around the $\Gamma$ point in the $t-t'$ model.

%\bibliography{refs}

\end{document}